\documentclass{aa}

\usepackage{graphicx}
\usepackage{txfonts}
\usepackage{soul}
\usepackage[usenames]{color}
\usepackage{setspace}

\begin{document} 

   \title{The role of cloud particle properties on WASP-39b transmission spectrum based on JWST/NIRSpec observations}

   \titlerunning{Cloud particle properties' role on transmission spectroscopy}
   \authorrunning{Roy-Perez et al.}

   \author{J. Roy-Perez
          \inst{1},
          S. Pérez-Hoyos
          \inst{1},
          N. Barrado-Izagirre
          \inst{1}
          \and
          H. Chen-Chen
          \inst{1}
          }

   \institute{Escuela de Ingeniería de Bilbao, Universidad del País Vasco (UPV/EHU), Bilbao, Spain\\
              \email{juan.roy@ehu.eus}
             }

   \date{Received March 27, 2024 / Accepted January 20, 2025}

  \abstract
   {Aerosols are capable of having a huge influence on reflected, emitted and transmitted planetary spectra, especially at wavelengths similar to their average sizes, but also extending to much longer and shorter wavelengths. The Near InfraRed Spectrograph (NIRSpec) using the PRISM mode on board of the James Webb Space Telescope (JWST) is providing valuable data of transit spectra over a wide spectral range that is able to cover the whole contribution of aerosols, potentially disentangling them from other constituents and thus allowing to constrain their properties.}
   {We aim to investigate whether NIRSpec/PRISM JWST transmission spectroscopy observations, in addition to being useful to detect and determine the abundance of gases more accurately than any previous instruments, are also capable of studying the physical properties of the aerosols in exoplanetary atmospheres.}
   {We perform nested sampling Bayesian retrievals with MultiNest library. We use the Planetary Spectrum Generator (PSG) and the Modelled Optical Properties of enSeMbles of Aerosol Particles (MOPSMAP) database as tools for the forward simulations and NIRSpec/PRISM JWST previously published observations of WASP-39b as input data.}
   {Retrievals indicate that models including an aerosol extinction weakly increasing or sharply decreasing with wavelength are decisively better than those with a flat transmission and that this increased degree of complexity is supported by the kind of data that JWST/NIRSpec can provide. Given other physical constraints from previous works, the scenario of weakly increasing particle extinction is favoured. We also find that this also has an effect on the retrieved gas abundances.}
   {JWST observations have the potential to study some physical characteristics of exoplanetary clouds, in particular their overall dependence of transmissivity with wavelength. In fact, it is important to implement more detailed aerosol models as their extinction may affect significantly retrieved molecular abundances.}

  \maketitle
%

\section{Introduction}\label{sec:Introduction}

    All the planets in the solar system  with a significant atmosphere have clouds and layers of aerosols \citep{Sanchez_Lavega_2011_Planetary_atmospheres_book}. These ensembles of particles play an important role in the atmospheric energy budget by absorbing the incoming sunlight and blocking the outgoing thermal emission \citep{Pater2015book}. But clouds are not only present in our closest neighbours, most of the hot Jupiters studied so far show a clear evidence of having clouds in their atmospheres \citep{Barstow2017, Sing2016, Tsiaras2018, Helling_2019, Barstow2020}. 

    Due to the viewing geometry, transmission spectra obtained during primary transits are very sensitive to the presence of clouds in the exoatmospheres \citep{Fortney2005}. Because of their opacity, clouds can block the stellar flux transmitted through the atmosphere and hide the spectral signals that would leave different molecules below them. Scattering effects caused by the aerosols can introduce slope-shaped effects in the observed spectra \citep{Kreidberg2018Exoplanets, Madhusudhan2019}.

    The discovery of the planet WASP-39b was reported in 2011 by \cite{Faedi2011}. It is a highly inflated Saturn-mass planet with a mass of $0.28 \, M_J$ and a radius of $1.27 \, R_J$ orbiting a G-type star. After its discovery, diverse studies considering different abundances of water, potassium and sodium  \citep{Sing2016, Heng2016, Fisher2016} tried to determine if the planet had a cloudy atmosphere, but the results were not conclusive. Further works tended to show a favourable tendency towards the cloudy atmosphere \citep{Barstow2017}, but the idea of WASP-39b having a cloudy atmosphere was generally accepted when \cite{Wakeford2018} completed the planet spectrum with the near-infrared (NIR) spectral range using data from the Wide Field Camera 3 (WFC3) onboard the Hubble Space Telescope (HST). Posterior analyses have continued treating WASP-39b as a cloudy planet trying also to constrain the properties of the clouds \citep{Tsiaras2018, Pinhas2018, Fisher2018, Pinhas2019, Kirk2019}.

    Parameterising all the processes involved with clouds using the minimum number of parameters is a challenging task. \cite{Barstow2020} presents a brief summary of the models used in the latest studies \citep{Barstow2017, Tsiaras2018, Fisher2018, Pinhas2019}. All these models are mainly focused on describing the vertical distribution of the cloud along the upper part of the atmosphere and determining at which pressure level it becomes opaque. They also tried to capture and to reproduce the wavelength dependence of the extinction efficiency of the cloud, but it is a difficult goal due to the limited spectral range of the observations. Thus, most of the models treat cloud extinctions as flat or Rayleigh shaped contributions.

    Here is where James Webb Space Telescope (JWST) starts playing a very important role in the study of clouds on exoplanets. Since its launch in December 2021 it has been observing transiting exoplanets with a much wider spectral range than any other previous instrument, such as those onboard HST or Spitzer Space Telescope. A wider range does not only capture more gaseous absorption bands from potentially a greater number of species, but it is also better constraining the smooth wavelength dependence of cloud opacity, as we will investigate here.

    In this work, we have used WASP-39b JWST observations already reduced and analysed in detail by previous works \citep{Rustamkulov2023} in order to demonstrate the possible contribution of more sophisticated cloud particle models. A recent work by \cite{Lueber2024} has investigated the same data using the aerosol parameterisation proposed by \cite{Kitzmann&Heng2018} but they were not able to disentangle aerosol properties from other contributions. Other works, such as \cite{Carone2023}, have shown that expected properties of the clouds and aerosols should be more complex than usually assumed. In this work, instead of computing aerosol distributions and properties from some ab initio hypotheses on the atmosphere, we have investigated such characteristics by treating them as free parameters, as is commonly done in many studies of solar system atmospheres. Our main goal is thus simply to try to determine an overall trend in the aerosol extinction, whether increasing, decreasing or constant with wavelength assuming the minimal number of a priori hypotheses over their nature and properties.
    
    To do so, we have structured this text as follows. In Sect. \ref{sec:Data} we present the main characteristics of the observational data. In Sect. \ref{sec:Methodology} we present the tools needed to perform the retrievals based on Bayesian inference and the modelisation of the atmosphere that we have been using. In Sect. \ref{sec:Results} we show the results obtained for the retrievals and in Sect. \ref{sec:Conclusions} we summary the conclusions that we have reached.

\section{Data} \label{sec:Data}

The spectral data were taken from the supporting material of \cite{Rustamkulov2023}. The observations were part of the JWST Transiting Exoplanet Community Early Release Science Program (ERS Program 1366). These data correspond to the 8.23 hours long transit of the planet WASP-39b from July 10, 2022, the first transit of the exoplanet observed with the Near InfraRed Spectrograph (NIRSpec) instrument of JWST using its PRISM mode. It covers the spectral range from 0.5 $\mu$m to 5.5 $\mu$m, visible to mid infrared, with a resolving power of 20-300.

In \cite{Rustamkulov2023} four different data reductions are presented to extract the transmission spectra, although they demonstrated that all of these lead to equivalent overall results. However, to guide their discussions throughout the paper they used the FIREFLy reduction \citep{RustamkulovFIREFLy}. Thus, for an easier comparison, we have decided to use also the FIREFLy reduction as the observational input data.

Due to the brightness of WASP-39, the exoplanet observations are saturated in the region from 0.8 $\mu$m to 2.3 $\mu$m, but data reduction alleviated this effect allowing to successfully measure fluxes in that region. However, as the original work warns, the 0.9 - 1.5 micron range calibration remains questionable.

\begin{figure}
   \centering
   \includegraphics[width=\hsize]{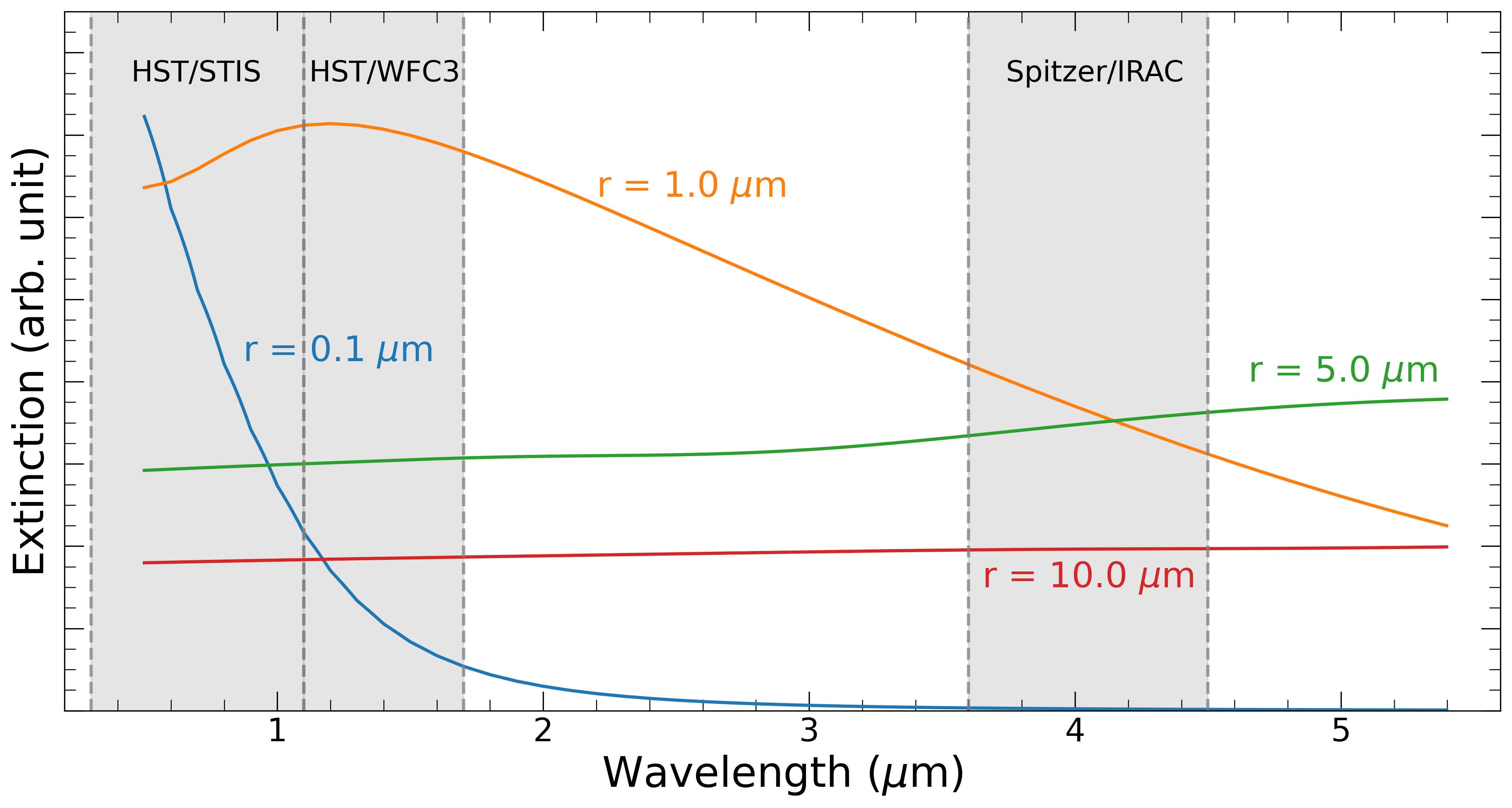}
   \caption{Simulated transmissivity with MOPSMAP for spherical aerosols of different radius all over the complete spectral range of JWST/NIRSpec-PRISM. HST/STIS, HST/WFC3 and Spitzer/IRAC spectral ranges are delimited by grey background for comparisons.}
   \label{Fig1}
\end{figure}

Prior to the JWST launch, there were only transmission spectra covering the wavelength range between 0.3 $\mu$m and 1.65 $\mu$m using the WFC3 and the Space Telescope Imaging Spectrograph (STIS) instruments of the Hubble Space Telescope. These observations were sometimes combined with wide photometric measurements from the Spitzer Space Telescope at 3.6 $\mu$m and 4.5 $\mu$m \citep{Barstow2017, Changeat_2022}. However, the selected NIRSpec/PRISM covers a wavelength range from 0.5 to 5.5 microns which, as we will argue in the following sections, has unique capabilities for constraining aerosol properties. Figure \ref{Fig1} sketches the difference in the spectral ranges covered by these three different telescopes.

\section{Methodology}\label{sec:Methodology}

   \subsection{Tools}\label{subsec:Tools}
   
   The forward simulations have been executed using the Planetary Spectrum Generator \citep{PSG2018}, PSG from now on. It is a radiative transfer suite developed by NASA's Goddard Space Flight Center team that allows the users to generate simulations of an extensive set of planetary bodies (both surfaces and atmospheres) over a wide spectral range. It is possible to simulate spectra for objects such as planets from the solar system, exoplanets or even comets. These simulated observations can be generated under the configuration of diverse instruments onboard space telescopes, terrestrial observatories or space missions. The user can modify a large variety of atmospheric and/or surface properties and parameters. It is also possible to define the geometry for the simulated observation: transiting exoplanets, nadir viewing or surface observations for landers, among others.

   Planetary spectra are generated using the core radiative-transfer model Planetary and Universal Model of Atmospheric Scattering \citep{PSGPUMAS, PSGPUMASScattering}, also known as PUMAS, which, using a one-dimensional and plane-parallel description of the atmosphere, computes high resolution spectra via line-by-line calculations and moderate resolution spectra using a correlated-k method. We use the line lists for $\rm{H_2O}$ \citep{Polansky2018}, $\rm{CO_2}$ \citep{Yurchenko2020}, $\rm{CO}$ \citep{Li2015}, $\rm{SO_2}$ \citep{Gordon2022}, $\rm{Na}$ \citep{Allard2019}, $\rm{H_2S}$ \citep{Gordon2022}, $\rm{CH_4}$ \citep{Yurchenko2024} and $\rm{K}$ \citep{Allard2016}. PUMAS takes into account the contribution of continuum processes such as Rayleigh scattering \citep{Maarten2005}, refraction, UV broad absorption and collision-induced absorption by $\rm{H_2-H_2}$ \citep{Abel2011} and $\rm{H_2-He}$ \citep{Abel2012}. The radiative-transfer model also includes a multiple scattering modelling.

    While the PSG online interface is very intuitive and didactic, it is also possible to install it locally through a Docker virtualisation system. Working locally with configuration files (with XML format) also allows to speed up the retrieval process by the use of the Application Program Interface (API) and the high-level programming language of your choice.

   PUMAS is able to include the contribution of aerosols, but PSG suite only includes information for a limited set of them. However, it is possible to add custom aerosol information via the configuration file including the data about three different parameters. The extinction cross section, $Q_{ext}$, is the attenuation coefficient normalised by the aerosol density and it characterises how effective the aerosol is in removing radiation from the incoming radiation beam. The single-scattering albedo, $\varpi_{0}$, determines the ratio of scattered to extinguished (absorbed plus scattered) light from the incident beam. And the asymmetry parameter, $g$, represents the phase function, i.e. the directional distribution of the scattered light. Alternatively, aerosol phase functions can be provided in a much more detailed way using Legendre polynomials, but given the kind of data we will be dealing with, this approach has not been considered in this work. Those quantities might be computed in different ways, using for example external databases as Modelled Optical Properties of enSeMbles of Aerosol Particles, referenced as MOPSMAP (see Sect. \ref{subsubsec:CloudModels}).

   As commonly done for atmospheric retrievals, we have employed a Bayesian approach \citep{Madhusudhan2011, Benneke2012, Lee2012, Line2013, Evans2016, Fisher2018}. We have used MultiNest \citep{FerozMultiNest, BuchnerPyMultiNest}, an inference tool based on Nested Sampling Monte Carlo algorithms \citep{Metropolis1953, Hastings1970, Skilling2006}. Open-sources codes as TauRex \citep{Waldmann2015} or Helios-r2 \citep{Kitzmann2020} already make use of MultiNest, but we have developed our own code to engage it to PSG.

   For each retrieval, MultiNest calculates the Bayesian evidence $Z$ of the model under study. This value is related to the probability that the model is actually representing the real observations regardless of the value of its parameters. The comparison of the Bayesian evidence given by different models will be the key to model selection. The Jeffrey's scale \citep{Jeffrey, BayesTrotta}, shown in table \ref{t1}, quantifies how much better one model is compared to another in terms of the Bayes factor. It is calculated as $\ln B_{1,2} = \ln Z_{1} - \ln Z_{2}$, where $Z_{1}$ and $Z_{2}$ are the Bayesian evidence values obtained for the models 1 and 2 respectively. When the Bayes factor is bigger than $5.0$ it is possible to state that one model is decisively better than the other. Nevertheless, when $\ln B_{1,2}$ is lower than 1 there is not enough evidence to claim which one is better. In this case the Occam's razor is followed favouring the model with a lower number of free parameters.

   \begin{table}[]
   \caption{Bayes factor values and their related evidences.} 
   \label{t1}
   \centering
   \begin{tabular}{cc}
   \hline\hline
   $\ln B_{1,2}$ & Evidence     \\ \hline
   0 - 1.0               & Inconclusive \\
   1.0 - 2.5            & Weak             \\
   2.5 - 5.0            & Moderate                  \\
   > 5.0            & Strong                           
   \end{tabular} 
   \tablefoot{More details in \cite{BayesTrotta}.}
   \end{table}

   MultiNest also returns the posterior probability distribution for each free parameter, thus allowing parameter determination. Following the law of large numbers, we assume the median of the marginalised probability distribution as the most likely value for each parameter. Similarly, uncertainties are calculated as 1$\sigma$ deviations from that value. This is really important because besides getting a result, we can affirm if we are sensible enough to a parameter and we are able to constrain a value with true physical meaning. In spite of the high dimensionality of the free parameter space, it is possible to visualise the correlation between parameters by plotting the marginalised probability distribution as a function of pair of parameters, in what is often referred to as a `corner plot'.

   \subsection{Atmospheric description}\label{subsec:AtmDesc}

   As stated before, PSG allows a huge number of options to define the atmosphere. In this subsection we will show which atmospheric parameters are assumed as fixed in our models and which are left as free parameters. The size of the star, $R_{*}$, determines the total amount of light reaching the planet, while the planetary diameter, $D_{pl}$, constrains the planetary surface blocking the star light. We allowed these parameters to vary but with the tight prior Gaussian distribution constraints calculated in \cite{Faedi2011}: $179,000 \pm 7,000$ km for $D_{pl}$ and $0.895 \pm 0.023 \, R_{\odot}$ for $R_{*}$.

   The scale height represents how compressed or extended the atmosphere of a planet is. It depends on the planetary gravity, the atmospheric mean molecular weight and its temperature \citep{Kreidberg2018Exoplanets}. We have assumed a Gaussian prior distribution for the planetary gravity of $log \, g = 0.63 \pm 0.05$ (with $g$ in $\mathrm{m} \, \mathrm{s}^{-2}$) based on \cite{Faedi2011}. Both parameters, $D_{pl}$ and $log \, g$, are referenced to the pressure reference level of 10 bar, which corresponds to the lowest layer of our model. The mean molecular weight will be calculated based on the assumed molecular abundances.

   \subsubsection{Vertical profiles}\label{subsubsec:VerticalProfiles}

   In this section we will discuss the vertical temperature and chemical species profiles. To describe them, we have defined a log-uniform pressure profile with 40 levels between $10^{-6}$ and $10^1$ bar.

   As explained in \cite{Line2015}, parameterising the temperature profile in planetary atmospheric retrievals is a challenging task. Different assumptions have been used for this purpose, see for example \cite{Guillot2010, Line2015} or \cite{Kitzmann2020}. We have implemented the parameterisation of \cite{Kitzmann2020} to determine the temperature profile to be included in our study. This parameterisation is given in the log-pressure space and takes up to three free parameters: the bottom temperature and two vertical gradients for separate atmospheric levels. As we do not expect having a thermal inversion in the atmsophere of WASP-39b \citep{Baxter2020} we have limited these slopes to be lower than $1.0$. As a first step in our study, we have performed three different retrievals, with one, two and three free parameters respectively to describe the temperature profile. The isothermal profile is better than the two parameter profile ($\ln B_{1,2} = 2.66$) and much better than the three parameter profile ($\ln B_{1,3} = 4.38$) so we will use this description for the rest of the paper.

   We have assumed an atmospheric temperature, $T_{iso}$, with a uniform prior distribution between 500 and 2000 K, although we expect to retrieve a value lower than the $T_{eq} = 1116 \, \pm \, 32$ K obtained by \cite{Faedi2011} since we will be most likely sensitive to stratospheric levels, which are expected to have temperatures lower than the equilibrium temperature \citep{Barstow2017}.

   The temperature profile is closely related to the molecular abundance profile of the atmosphere. We have defined the planet as mainly composed of $\rm{H_2}$ and $\rm{He}$, and at first instance we consider the presence of $\rm{H_2O}$, $\rm{CO_2}$ $\rm{CO}$, $\rm{SO_2}$ and $\rm{Na}$ in the upper atmosphere, the chemical species with significant detections in \cite{Rustamkulov2023}. Performing retrievals where the presence of other molecular abundances were individually included, we concluded that $\rm{H_2S}$ ($\ln B_{\rm{H_2S}} = 2.71$) must also be included in the simulations. Nevertheless $\rm{CH_4}$ ($\ln B_{\rm{CH_4}} = 0.37$) and $\rm{K}$ ($\ln B_{\rm{K}} = -0.60$) did not show enough evidence to be taken into account. We consider a vertically uniform distribution for the molecular components. Fixing the abundance of hydrogen and helium, we let the volume mixing ratio of the rest of the abundances vary during retrievals with uniform prior distribution between $10^{-12}$ and $10^{-1}$.

   \subsubsection{Cloud vertical profiles}\label{subsubsec:CloudPVModels}

   As mentioned in the introduction, the cloud vertical distribution along exoplanetary atmospheres plays an important role in the transit spectroscopy given that it can determine at which height the atmosphere becomes opaque. As a first step in our study, we have carried out a model selection to determine which cloud vertical profile explains better the observations. Once we have an indication of the best vertical cloud model, we will use it as a reference for the cloud extinction investigation.

   We have studied the most common vertical distributions in the literature: A. A uniform vertical distribution \citep{Fisher2018}; B. a step shaped opaque grey cloud distribution \citep{Kreidberg2014}; C. a step shaped distribution \citep{Benneke2015, Barstow2017}; D. a bottom opaque grey cloud combined with an upper uniform cloud \citep{Tsiaras2018, Pinhas2019} and E. a slab parameterisation \citep{Barstow2017}. A graphic display of these vertical parameterisations can be found in Fig. 1 of \cite{Barstow2020}.

   Models A and B are the simplest ones used in our simulations, each one including only a single free parameter: the aerosol abundance and the pressure-top level of the optically thick cloud, respectively. While it is not physically realistic to expect a uniform vertical distribution, this is justified  by the fact that our simulations are only being sensitive to a relatively narrow region of the atmosphere of around one or two decades in pressure.
   
   A second degree of complexity is achieved with models C and D, which involve two free parameters. The step model C defines a uniform abundance for the lowest part of the atmosphere below a given pressure level and then no aerosols at all above that level. Model D defines the pressure-top level of the optically thick cloud and the aerosol abundance of the extended cloud above that same level.

   Finally, model E defines a uniform distribution between two different pressure levels with no aerosol loading at the rest of the layers thus requiring three free parameters.

   \subsubsection{Cloud extinction models}\label{subsubsec:CloudModels}

   The effect of aerosols in our model is included using the three parameters presented in Sect. \ref{subsec:Tools}. Generally speaking, all of them depend on wavelength, which increases enormously the degeneracy of the problem. However, given the peculiar geometry for transmission spectroscopy, we find that $\varpi_{0}$ and $g$ do not produce any significant variation in the simulations, in agreement with \cite{Barstow2017}, and allow us to focus on $Q_{ext}$ and its dependence on wavelength to compare cloud models.

   Due to the relatively narrow wavelength range covered by observations prior to the JWST, most of the works on exoplanet atmospheres have assumed a flat or a Rayleigh-shaped (scattering efficiency scaling as $1 / \lambda^4$) cloud transmissivity \citep{Barstow2017}. This can be understood from Fig. \ref{Fig1}, which shows the simulated extinction over the entire spectral range of NIRSpec-PRISM for different spherical aerosol sizes. It can be seen how the transmissivity for each aerosol tends to peak at the wavelengths similar to their size, a conclusion that holds in general terms even for different particle shapes. Along HST spectral ranges, delimited by a grey background, small-sized aerosols match with an extinction similar to a Rayleigh scattering and large-sized ones could be approximated with flat grey clouds. Nevertheless, this assumption cannot be extrapolated to the JWST spectral range where the $1 / \lambda^4$ behaviour disappears for the small aerosols and where a flat transmissivity would underestimate the transmissivity at longer wavelengths.

   We will use as a reference the flat model (model I), where the value of $Q_{ext}$ is constant with wavelength. As mentioned in the introduction, it is the simplest model that we could define and it is commonly used when studying the atmospheres of exoplanets. In order to compare with the flat cloud model, we have defined a number of competing scenarios, each one with a different dependence of $Q_{ext}$ with wavelength.

   Model II is the free slope model, in which we parameterise $Q_{ext}$ as linearly dependent on wavelength leaving the slope, $m_{Cl}$, as a free parameter. Although there is no direct physical interpretation for this model, positive slope values would resemble larger particles while negative slopes would mimic the behaviour of smaller particles. Thus, model II, including only one extra free parameter, is a simple first approach to model particle sizes.

   Model III is based on the $\textup{\r{A}}$ngström exponent. \cite{Angstrom} proposed that the optical thickness of an aerosol would behave with the wavelength of light as

   \begin{equation}\label{eq:5}
       \frac{\tau_{\lambda}}{\tau_0} = \left( \frac{\lambda}{\lambda_0} \right) ^{-\alpha},
   \end{equation}
   where $\alpha$ (always positive) is defined as the $\textup{\r{A}}$ngström exponent. In this case the parameter is inversely proportional to the particle size: the smaller the value of $\alpha$, the bigger the aerosol. This cloud opacity dependence with wavelength has already been used in exoplanet atmospheric studies as \cite{Lecavelier2008, McDonald2017, Pinhas2019} or \cite{Barstow2020}.
   
   Given that the optical thickness and the extinction cross section are directly related, we can retrieve the best value for the $\textup{\r{A}}$ngström parameter and build again the values of $Q_{ext}(\lambda)$ using the value of reference by

   \begin{equation}\label{eq:6}
       Q_{ext, \lambda} = Q_{ext,ref}\left( \frac{\lambda}{\lambda_{ref}} \right) ^{-\alpha}.
   \end{equation}
   
   As it is possible that some aerosol distributions produce an opacity growing with wavelength when covering the whole JWST spectral range (see Fig.\ref{Fig1}) we have decided to let the parameter $\alpha$ also reach negative values that would result in opacity dependences similar to those produced by large aerosols. With this parameterisation we are also including a steeper dependence with wavelength than the simple linear model discussed above.

   A more realistic approach is achieved with cloud extinction model IV, which is generated using MOPSMAP \citep{MOPSMAP}, a database of optical properties of aerosol ensembles. The data set can provide the optical properties of single particles (or narrow size bins) on a grid of sizes, shapes, and refractive indices. 

   MOPSMAP allows the user to define the particles as spheres, spheroids, irregular bodies or even a mixture of them. Aerosols present in exoplanet atmospheres must also have different shapes, but the latest literature shows that just determining particle sizes and composition is already a difficult issue \citep{Kitzmann&Heng2018}. Thus, for simplicity, we have defined the aerosols in our simulations with a spherical shape, as done for example in \cite{Sushuang2023} and \cite{Lueber2024}. Please note that such spherical shapes would leave a fingerprint in direct imaging observations at different phase angles and could potentially lead to the confirmation of spherical droplets the same way it has been done for Venus \citep{Garcia2014}. 

   Whenever necessary, we will use for the aerosols a log-normal size distribution \citep{Hansen}, commonly used in solar system atmospheric studies \citep{BarstowVenus2012, Arteaga2023, Toledo2019} and also used for brown dwarfs and exoplanets \citep{Ackerman2001, Benneke2015}. The amount of particles for a given size $r$ is then described by:

   \begin{equation}\label{eq:1}
       n \left( r \right) = \frac{1}{\sqrt{2 \pi}} \frac{1}{\sigma_g r}  \exp{ \left[-\frac{1}{2} \left( \frac{\ln{r} - \ln{r_g}}{\sigma_g} \right)^2 \right]},
   \end{equation}
   where $\ln{r_g}$ and $\sigma_g$ are respectively the mean and the standard deviation of the distribution in logarithmic scale. To describe this distribution, however, we use the effective radius, $r_{eff}$, and the effective variance, $v_{eff}$, which are more intuitive when working with aerosol extinction since they are the mean and deviation of a size distribution weighted by the cross section of the particles \citep{Hansen}. They are related to the log-normal distribution parameters according to the following equations:
   
   \begin{equation}\label{eq:2}
       r_g = \frac{r_{eff}}{\left(1+v_{eff} \right)^{\frac{5}{2}}},
   \end{equation}
   
   \begin{equation}\label{eq:3}
       \sigma_g^2 =  \ln{\left(1+v_{eff} \right)}.
   \end{equation}

   Concerning the spectral properties of aerosols, MOPSMAP needs both the real and the imaginary part of the refractive index vector. This parameter is determined by the composition of the aerosols, with the real part describing the refraction of the particles and the imaginary part describing their absorption \citep{Hansen, Liou}.

   We compute $Q_{ext}(\lambda)$ data for different $r_{eff}$ particle distributions fixing $r_{\sigma}$ to $0.1$. Given that we do not have any previous information about the composition or structure of the aerosols in WASP-39b, at the first instance we decided to fix both the real and the imaginary part of the refractive index at every wavelength to $n_r = 1.4$ and $n_i = 0.0001$. We will discuss the potential role of the imaginary refractive index in the retrievals  later in Section \ref{sec:Results}.

   There is an underlying assumption that must be taken with care: that the bayesian evidences for these different parameterisations are comparable with each other. Given that all of them only include one extra free parameter with respect to model I, and that the volumes of the parameter hyper-space are similar, we assume that it is possible to compare the evidence as in other cases. Table \ref{tprioris} summarises the prior probability distributions introduced in MultiNest for every free parameter included in the simulations.

    \begin{table}[]
    \caption{Summary of the prior probability distributions for the free parameters included in the retrievals.}
    \label{tprioris}
    \centering
    \begin{spacing}{1.2}
    \begin{tabular}{lccc}
    \hline\hline
    Parameter & Distribution & Range & Units       \\ \hline
    
    $D_{pl}$    & Gaussian    & $179,000 \pm 7,000$   & km      \\
    $R_{*}$     & Gaussian    & $0.895 \pm 0.023$   &   $R_{\odot}$ \\
    $log \, g$  & Gaussian    & $0.63 \pm 0.05$     & $\mathrm{m} \, \mathrm{s}^{-2}$ \\
    $T_{iso}$  & Uniform   & [$500.0$, $2000.0$]   & K      \\ 
    $X_i$  & Log-uniform   & [$10^{-15}$, $10^{-1}$]   & -     \\ \hline
    \textit{Cloud profile} & & & \\
    Cloud   & Log-uniform     & [$10^{-15}$, $10^{-1}$]     & $\mathrm{kg/kg}$      \\
    $P_{Top}$  & Log-uniform      &  [$10^{-6}$, $10^{1}$]  & bar  \\ 
    $P_{Bot}$  & Log-uniform      &  [$10^{-6}$, $10^{1}$]  & bar  \\ \hline
    \textit{Cloud extinction} & & & \\
    $m_{Cl}$ & Uniform  & [$-2.0$, $2.0$]  &  - \\
    $\alpha$ & Uniform  & [$-5.0$, $5.0$]  &  - \\
    $r_{eff}$ & Log-uniform  & [$0.005$, $30.0$]  &  $\mu$m

    \end{tabular} 
    \tablefoot{$X_i$ represents the volume mixing ratio abundance of the molecule $i$.}
    \end{spacing}
    \end{table}

\section{Results and discussion}\label{sec:Results}

This section is organised as follows. We first discuss the results for determining the vertical cloud profile that agrees better with the observations. Then, using the best retrieved vertical profile, we perform the aerosol extinction model selection. Once we have a full description of the most plausible aerosol spectroscopic behaviour we analyse how these assumptions affect the rest of free parameters of the retrievals. The idea is to bracket the retrievals with a range of possible assumptions to check the dependence of the obtained results with the description of the clouds and to determine their robustness. We lastly study the possibility to include molecular abundances initially discarded when using grey flat cloud extinctions and study the sensitivity to other parameters, not initially considered, which may have an effect on the previous analyses.

\subsection{Vertical cloud profile model selection}\label{subsec:VCModelSelection}

   Table \ref{t2} shows the Bayes factor obtained for each cloud vertical distribution model using the uniform model as reference, the simplest one. Based on Jeffrey's scale (Table \ref{t1}), all cloud models seem quite similar in evidence with the exception of model B, the opaque step, which is clearly disfavoured by the data. Thus, given that model A gives same evidence that the rest of models but using the minimum number of free parameters, following Occam's razor we conclude that the uniform vertical distribution is the best choice for describing the observations.

   \begin{table}[]
    \caption{Logarithm of the  Bayesian evidence obtained for the cloud vertical models and the relative Bayesian factor values when comparing to the most likely model.}
    \label{t2}
    \centering
    \begin{tabular}{llc}
    \hline\hline
     & Model  & $\ln B_{1,2}$ \\ \hline
    A & Uniform        & ...    \\
    B & Opaque Step          & -3.87    \\
    C & Step      & -0.39    \\
    D & Opaque Step + Uniform    & -0.01 \\ 
    E & Slab      & -0.37
    
    \end{tabular} 
    \end{table}

    Inspecting retrieval results we realize that all vertical models have evolved towards a similar solution. We find that all vertical models, except model B, result in a distribution of particles that reaches $\tau = 1$ in nadir geometry at levels between 100 mbar and 1 bar. At this level the atmosphere has already become opaque and all the incoming stellar light is being blocked. Model B, instead, does not produce enough extinction at the upper levels, so it compensates for this lack of opacity placing the cloud top at higher atmospheric levels. Anyway, as we have just stated, we henceforth proceed using the uniform cloud vertical profile.

\subsection{Cloud extinction model selection}\label{subsec:AEModelSelection}

\begin{figure*}
   \centering
   \includegraphics[width=\hsize]{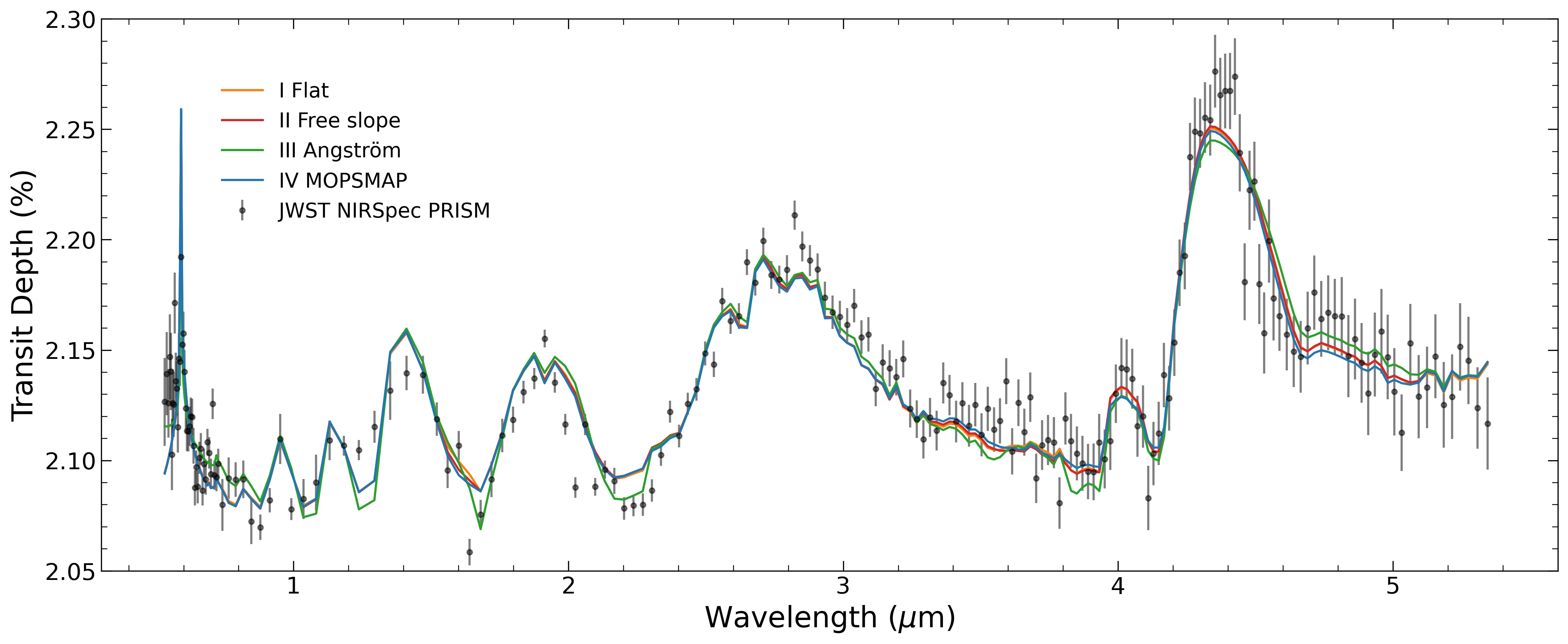}
   \caption{Best fits for the cloud extinction models once their retrievals are finished. Observational data are shown as black/grey dots with error bars.}
              \label{FigEspecs}
    \end{figure*}

Table \ref{t3} shows the Bayes factor obtained for each cloud extinction model compared to the flat model and Figure \ref{FigEspecs} shows the best fits to the observational data achieved in each retrieval. It is important to remark that, even if all the different models reach a good fit with the observations, barely distinguishable between them, there is a clear difference in the retrieved evidence meaning that the particle extinction model also plays a substantial role that can be constrained by the JWST observations.

\begin{table}[]
\caption{Bayesian factor values obtained for the cloud extinction models when comparing to the flat model.}
\label{t3}
\centering
\begin{tabular}{llc}
\hline\hline
 & Model  & $\ln B_{1,2}$ \\ \hline
I  &  Flat        & ...    \\
II & Free slope          & -1.40    \\
III & $\textup{\r{A}}$ngström      & 8.02    \\
IV & MOPSMAP      & 5.57
    
\end{tabular} 
\end{table}

\begin{figure}
   \centering
   \includegraphics[width=\hsize]{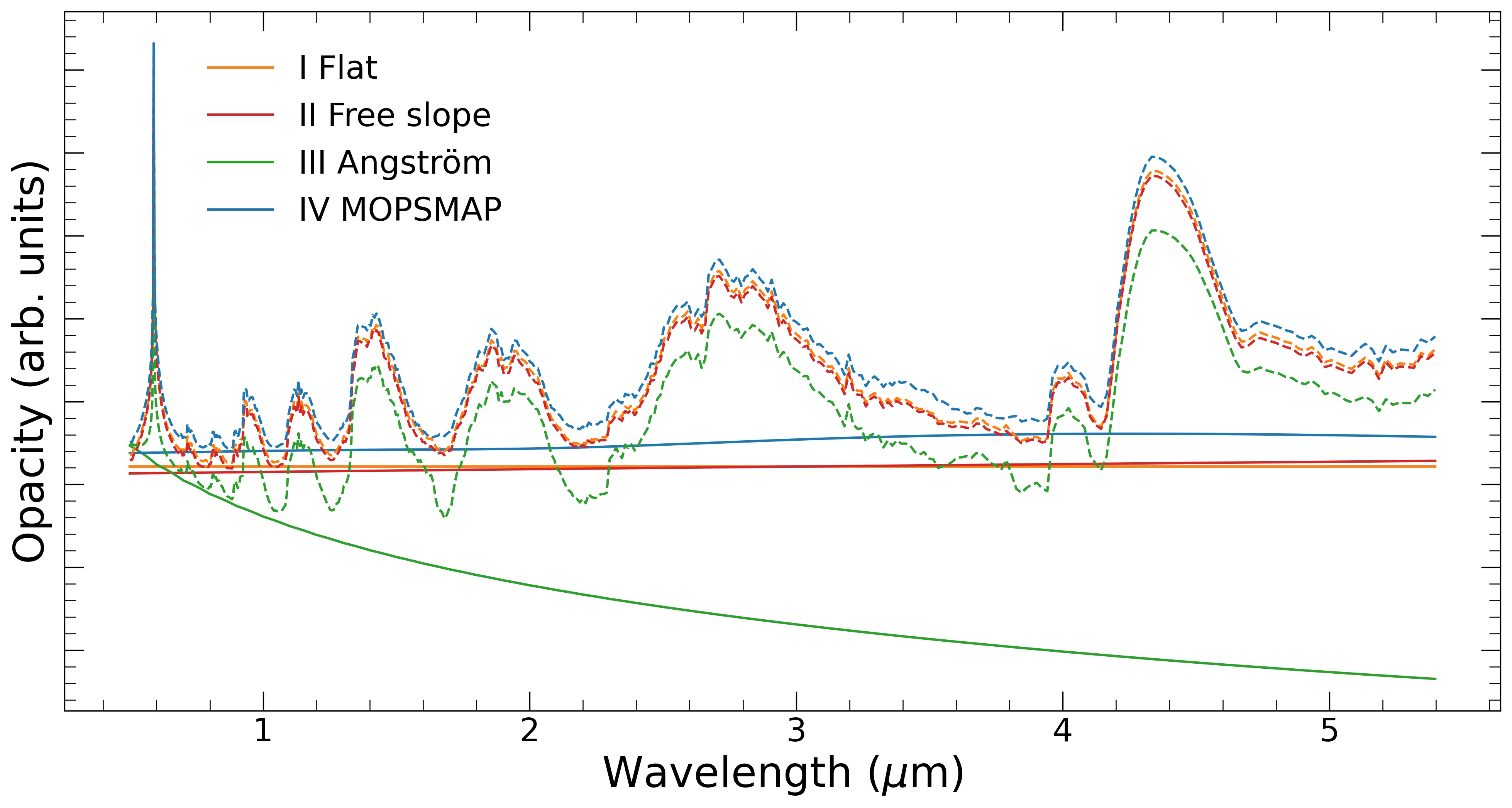}
   \caption{Cloud opacity retrieved for each cloud extinction model in comparison to their total atmospheric opacity.}
              \label{FigOpacities}
    \end{figure}

In base to Table \ref{t1} we could argue that Model III has been the best choice for describing the aerosol opacity. As we can see in Fig. \ref{FigOpacities}, which shows the retrieved aerosol opacities for each model, the $\textup{\r{A}}$ngström model has evolved towards producing high optical depths at shorter wavelengths and a decrease towards longer wavelengths. For $1.0$ $\mu$m it reaches values of $\tau = 1$ at levels similarly to the results for the flat extinction cloud, but at $4.0$ $\mu$m, at the same pressure levels, it only reaches values of $\tau = 10^{-2}$ - $10^{-3}$. Nevertheless, in Fig. \ref{FigOpacities} we can observe a clear dichotomy. Models II and IV, not preferred in terms of Bayesian evidence, have not been able to reach that heavy opacity decrease generating instead opacities with a slight increase with wavelength in comparison to Model I.

We suggest thus two possible aerosol opacity contributions when fitting Wasp-39b transit spectra. On the one hand, the opacity heavily decreasing with wavelength, resembling the effect of small particles. As we will see later, this aerosol model requires much higher molecular abundances to compensate for the thin cloud effect at longer wavelengths and this implies also very high values for the mean molecular weight of the atmosphere of $\bar{\mu}_{atm} = 3.50$ g $\mathrm{mol}^{-1}$. On the other hand, the slightly increase of opacity resembling the effect of big aerosol particles results in smaller molecular abundances with mean molecular weights more in line with  those expected for the planet ($\bar{\mu}_{atm} \approx 2.60$ g $\mathrm{mol}^{-1}$).

\subsection{Parametric study}\label{subsec:ParametricStudy}

Figures \ref{FigCornerPlot} and \ref{FigCornerPlot2} show the corner plots obtained for Model III and Model IV respectively, each of them being the most representative models of both cloud extinction behaviours already presented. The algorithm has been sensitive enough to all the parameters, achieving Gaussian-like posterior distributions in all cases, maybe with the exception of log $\rm{H_2S}$ for model IV (see Fig. \ref{FigCornerPlot2}). This means that the retrieval has always been capable to determine the most probable parameter value and its corresponding uncertainties. From both figures we can see how, as previously claimed by \cite{Fisher2018} and recently demonstrated in \cite{Lueber2024}, the spectral resolution of JWST observations has enough potential to break the normalization degeneracy. Retrievals performed with these observations show no correlation between the molecular abundances and the reference planetary diameter. $D_{pl}$ shows only a strong correlation with the radius of the central star.

Model III shows a strong anticorrelation between the aerosol abundance and the parameter $\alpha$. This is due to the strong requirement of the model to have a substantial particle extinction below 1 micron. In Model IV, instead, there is a slight correlation between molecular and cloud abundances, as the thicker the cloud, the more gaseous absorption the model requires to fit the observed molecular bands. Lastly, it is possible to appreciate in both corner plots the correlation between the retrieved volume mixing ratios of $\rm{H_2O}$ and $\rm{CO_2}$, also noted in \cite{Lueber2024}, due to their competing effect for shaping the peak around $2.8$ $\mu$m.

Retrieved parameter values from the posterior distributions for models III and IV are summarized in Table \ref{t4}. As expected, $D_{pl}$ and $R_{*}$ are compatible with the prior distribution values included from \cite{Faedi2011}. However, there is a discrepancy in the obtained value of $log \, g$, it does not fit with the value given by \cite{Faedi2011} nor with the one determined by \cite{Mancini2018}. Obtaining the same discrepancy, \cite{Lueber2024} proposed that this enhanced value can be due to the retrieval trying to obtain a diminished scale height. As anticipated in section \ref{sec:Methodology}, the retrieved $T_{iso}$ value is somewhat lower than the $T_{eq}$ of \cite{Faedi2011}.

\begin{table}[]
\caption{Parameter a posteriori values and uncertainties.}
\label{t4}
\centering
\begin{spacing}{1.4}
\begin{tabular}{lcc}
\hline\hline
Parameter &  Posterior Model III  & Posterior Model IV    \\ \hline

$D_{pl}$ (km)       & $175,000^{+2,000}_{-3,000}$  & $175,000^{+3,000}_{-3,000}$         \\
$R_{*}$  ($R_{\odot}$)    & $0.906^{+0.011}_{-0.015}$   & $0.909^{+0.013}_{-0.016}$      \\ 
$log \, g$ ($\mathrm{m} \, \mathrm{s}^{-2}$)    & $0.72^{+0.03}_{-0.04}$  & $0.73^{+0.03}_{-0.04}$       \\ 
$T_{iso}$ (K)    & $960^{+50}_{-40}$    & $870^{+40}_{-30}$     \\ 
log Cloud        & $-8.83^{+0.33}_{-0.36}$    & $-6.58^{+0.19}_{-0.16}$        \\
log $\rm{H_2O}$   & $-2.50^{+0.11}_{-0.10}$   & $-2.73^{+0.16}_{-0.15}$       \\
log $\rm{CO_2}$  & $-3.70^{+0.13}_{-0.12}$   & $-3.97^{+0.18}_{-0.18}$        \\
log $\rm{CO}$  & $-1.41^{+0.15}_{-0.16}$   & $-1.99^{+0.33}_{-0.32}$       \\
log $\rm{SO_2}$  & $-4.70^{+0.19}_{-0.17}$   & $-5.05^{+0.23}_{-0.21}$       \\
log $\rm{Na}$  & $-5.19^{+0.16}_{-0.16}$  & $-5.14^{+0.21}_{-0.19}$  \\
log $\rm{H_2S}$  & $-2.23^{+0.15}_{-0.14}$   & $-6.29^{+2.57}_{-3.83}$       \\
$\alpha$  & $2.89^{+0.57}_{-0.51}$ & - \\
log $r_{eff}$  & - &  $0.55^{+0.03}_{-0.03}$

\end{tabular} 
\tablefoot{Retrieved values are the median of the marginalised probability distribution of each parameter in linear scale, and their uncertainties are calculated taking into account the 1$\sigma$ deviations from this value.}
\end{spacing}
\end{table}

Regarding the retrieved chemical abundances, we make a comparison with recent works that have tried to constrain or at least detect different molecular abundances using different instrument observations. For the Model III retrieval we have obtained a value of log $\rm{H_2O} = -2.50^{+0.11}_{-0.10}$ with NIRSpec PRISM observations, compatible with the value ranging from $-3$ to $-2.5$ derived from the studies of \cite{Rustamkulov2023} employing the same instrument observations. Our water abundance calculated is also compatible with the values ranging from $-3.3$ to $-1.4$ derived from \cite{Ahrer2023ERS} studies with NIRCam observations. And it is even consistent with the results (log $\rm{H_2O} = -2.4$ to $-1.2$) derived from \cite{Powell2024} using near-infrared JWST observations. Nevertheless, we obtain a slightly lower value than those derived from \cite{Fisher2024} and \cite{Feinstein2023ERS} (log $\rm{H_2O} = -2$ to $-1$ and log $\rm{H_2O} = -1.86$ respectively), both employing Near Infrared Imager and Slitless Spectrograph (NIRISS) observations. A higher value of log $\rm{H_2O} = -1.37$ was also obtained by \cite{Wakeford2018} employing a combination of HST/WFC3, HST/STIS, Very Large Telescope (VLT), and Spitzer data. And our $\rm{H_2O}$ value is higher than those retrieved in \cite{Lueber2024} with the different JWST instrument observations, higher than the log $\rm{H_2O} = -4.85$ to $-3.14$ value retrieved with PRISM data in \cite{Constantinou2023} and higher than the value log $\rm{H_2O} = -4.07^{+0.72}_{-0.78}$ derived from HST and Spitzer in \cite{Pinhas2019}. As explained in \cite{Fisher2024}, these spread results are exposing the difficulty of measuring atmospheric chemical abundances, even with JWST observations.

From the strong footprints left by $\rm{CO_2}$ in the transit spectrum at $2.8$ $\mu$m and $4.3$ $\mu$m we have retrieved an abundance of log $\rm{CO_2} = -3.70^{+0.13}_{-0.12}$. This value is higher than the value of log $\rm{CO_2} \approx -5.0$ retrieved from \cite{Ahrer2023ERS} for atmospheric levels close to 1 bar, but remains below the calculated value log $\rm{CO_2} = -2.72^{+0.31}_{-0.51}$ using NIRISS observations in \cite{Fisher2024}. For the carbon monoxide we obtain a value log $\rm{CO} = -1.41^{+0.15}_{-0.16}$, higher than the values given by \cite{Grant2023} and \cite{Lueber2024} and the value derived from Extended Data Fig. 2 of \cite{Ahrer2023ERS} (log $\rm{CO} = -2.63$, log $\rm{CO} = -2.85^{+0.17}_{-0.28}$ and log $\rm{CO}$ ranging from $-2$ to $-3$, respectively).

For the $\rm{SO_2}$, the expected chemical specie producing the peak at $4$ $\mu$m, we obtain a volume mixing ratio of log $\rm{SO_2} = -4.70^{+0.19}_{-0.17}$. \cite{Rustamkulov2023} suggested values between -5 and -6 from NIRSpec PRISM observations, and \cite{Alderson2023} gave a value of log $\rm{SO_2} = -5.6^{+0.1}_{-0.1}$. And from near-infrared JWST observations, \cite{Powell2024} proposes an abundance from $-4.6$ to $-6.3$. For the sodium we are getting log $\rm{Na} = -5.19^{+0.16}_{-0.16}$, compatible with the value of log $\rm{Na} = -5.15^{+0.44}_{-0.35}$ obtained by \cite{Lueber2024} for its non-grey PRISM study. And for the $\rm{H_2S}$ we obtain log $\rm{H_2S} = -2.23^{+0.15}_{-0.14}$, an enhanced value when comparing with the value obtained by \cite{Lueber2024} ($\rm{H_2S} = -4.75^{+0.42}_{-0.38}$) and the derived value from Extended Data Fig. 2 of \cite{Ahrer2023ERS} ($\rm{H_2S} \approx -3.5$).

Results from model III are substantially higher than any other previous work, which makes sense, as none of them used such an extreme wavelength dependence with wavelength for the particle extinction. As already stated, such an enhanced molecular abundance would result in a huge molecular weight of $\bar{\mu}_{atm} = 3.50$ g $\mathrm{mol}^{-1}$, which is probably unphysical.

\begin{figure*}
   \centering
   \includegraphics[width=\hsize]{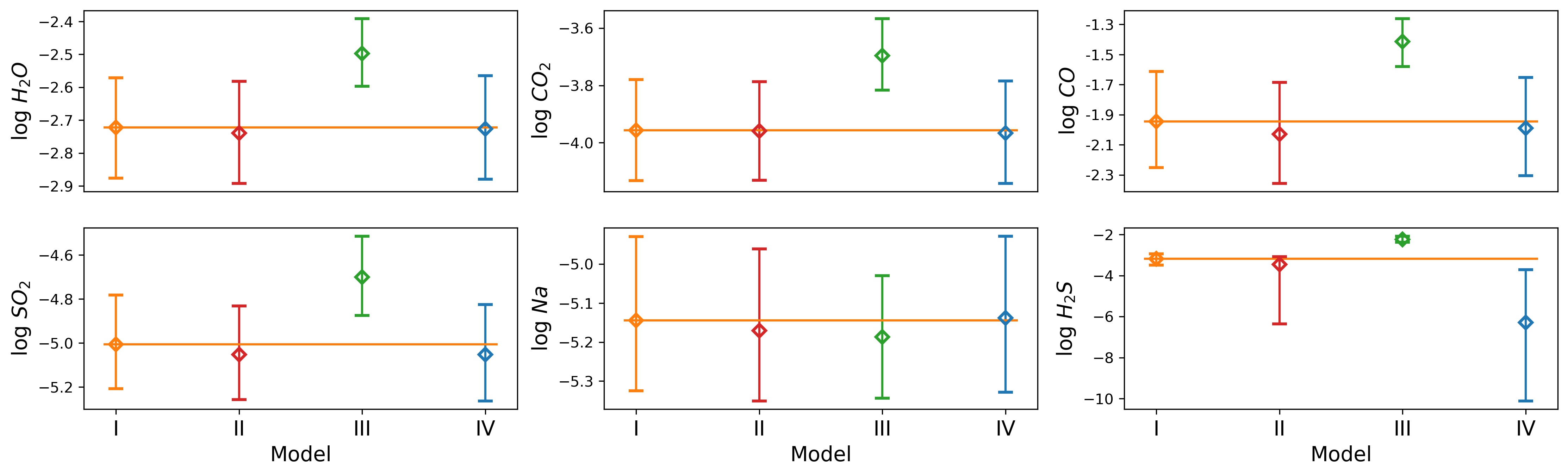}
   \caption{Comparison of the retrieved chemical abundances for the different cloud extinction models: Flat model (I), Free slope model (II), $\textup{\r{A}}$ngström model (III) and MOPSMAP model (IV). The horizontal line shows the value obtained for model I as a visual reference.}
              \label{FigResults}
    \end{figure*}

However, the focus of this section is on the bias that the different cloud extinction models may have on the retrieved molecular abundances, specially when comparing to the Flat Model. In Fig \ref{FigResults} we show the retrieved molecular abundances for the four different cloud extinction models. The first thing that we can observe is how Model III retrieved abundances are always higher, except for the sodium, than those for the other models. As previously explained, it is due to spectral shape of the retrieved cloud optical thickness. All the clouds reach similar $\tau$ values at shorter wavelengths, thus similar $\rm{Na}$ opacities are required, but Model III cloud becomes optically thin at higher wavelengths while the other clouds continue being opaque at those wavelengths. That lack of opacity is thus compensated by enhanced molecular abundances. While probably unrealistic, this is a clear warning that, if our cloud extinction model is far from the real cloud, then our chemical abundances may be off by orders of magnitude.

The volume mixing ratios obtained for Model IV are lower than those retrieved for Model III, producing a mean molecular weight similar to the Jovian value. These results fit better with the bibliographic values for $\rm{CO}$, $\rm{SO_2}$ and $\rm{H_2S}$ previously stated. In general, if clouds were composed of larger particles, using a flat cloud extinction modelisation for the atmospheric retrievals could be a good approximation, but it could be overestimating some molecular abundances. For Wasp-39b we have obtained with Model II and Model IV cloud optical depth that increases slightly with wavelength, what could mask and compensate the effect that $\rm{H_2S}$ would leave in the transmission spectrum, requiring thus a lower abundance.

As the different cloud extinction models are masking the required molecular opacities to fit the spectral observations, one could wonder if it could even compromise the detection of those chemical species. To find out, we have repeated Model III and Model IV retrievals but excluding the presence of $\rm{CO}$, $\rm{SO_2}$ and $\rm{H_2S}$. Bayes factors obtained for each gas removal retrieval when comparing to the complete model retrieval are shown in table \ref{t6}. Model III presents always strong evidence to include all the chemical abundances. Nevertheless, Model IV presents strong evidence to include $\rm{CO}$ and $\rm{SO_2}$, but not for $\rm{H_2S}$. Following Occam's Razor, the value $\ln B = -0.86$ is not enough to confirm the presence of that molecule in the atmosphere of the planet. We speculate that the slight increase with wavelength of the cloud opacity is masking the print that $\rm{H_2S}$ would leave in the spectrum. It must be taken into account that these numbers consider the whole spectral range at the same time, not particular spectral windows, so this reasoning cannot substitute other detection processes which are much more sensitive.

\begin{table}[]
\caption{Bayesian factor values obtained for cloud extinction models III and IV when excluding different chemical species in comparison to the original retrieval value.}
\label{t6}
\centering
\begin{tabular}{lcc}
\hline\hline
Gas Removal  & Model III  & Model IV  \\ \hline
$\rm{CO}$        & -18.83  &  -12.45  \\
$\rm{SO_2}$         & -17.64 &  -10.98 \\
$\rm{H_2S}$      & -14.00  & -0.86
    
\end{tabular} 
\end{table}

\subsection{Sensitivity to other parameters}\label{subsec:FurtherParams}

Once the best cloud model, in the sense of the one with the highest evidence supporting it, was established, we tried to make further parametric sensitivity studies. We will focus here in two main aspects. First, we will try to include methane, as it has been shown to be an elusive component \citep{Rustamkulov2023}. The optically thin spectral behaviour of cloud extinction Model III could possibly be partially compensated by extra molecular opacities, as for example $\rm{CH_4}$. Second, we will increase the complexity of our cloud particles description, by means of adding or removing light absorption and check if it affects the retrieved aerosol spectral behaviour. The first exploration implies adding the presence of $\rm{CH_4}$ to Model III and Model IV, as we did previously in Sect. \ref{sec:Methodology}. For the second, we will keep using in Model IV a constant in wavelength imaginary refractive index, $n_i$,  but using its value as a new free parameter.

Regarding the inclusion of methane we obtain a Bayes factor $\ln B_{1,2}=-0.52$ when trying to incorporate it in Model IV. This implies that the inclusion of $\rm{CH_4}$ to this model is not statistically supported. The upper limit retrieved for methane abundance is similar to those calculated in works as \cite{Rustamkulov2023, Ahrer2023ERS} or \cite{Fisher2024}. Nevertheless, the calculated $\ln B_{1,2}=8.66$ shows that there is evidence enough to include $\rm{CH_4}$ in our simulations when cloud extinction Model III is implemented. In this case the algorithm has retrieved a methane abundance of log $\rm{CH_4} = -5.85^{+0.14}_{-0.14}$, which also coincides with the upper limits of the works stated before.

On the other hand, the Bayes factor $\ln B_{1,2}=-0.56$ calculated when including $n_i$ as a free parameter to cloud extinction model IV shows that there is not evidence enough to be sensitive to the absorption of the aerosol. Posterior distributions for this retrieval show that any value of $n_i$ smaller than $10^{-2}$ is enough for reaching a good fit with the observations. 

\begin{figure}
   \centering
   \includegraphics[width=\hsize]{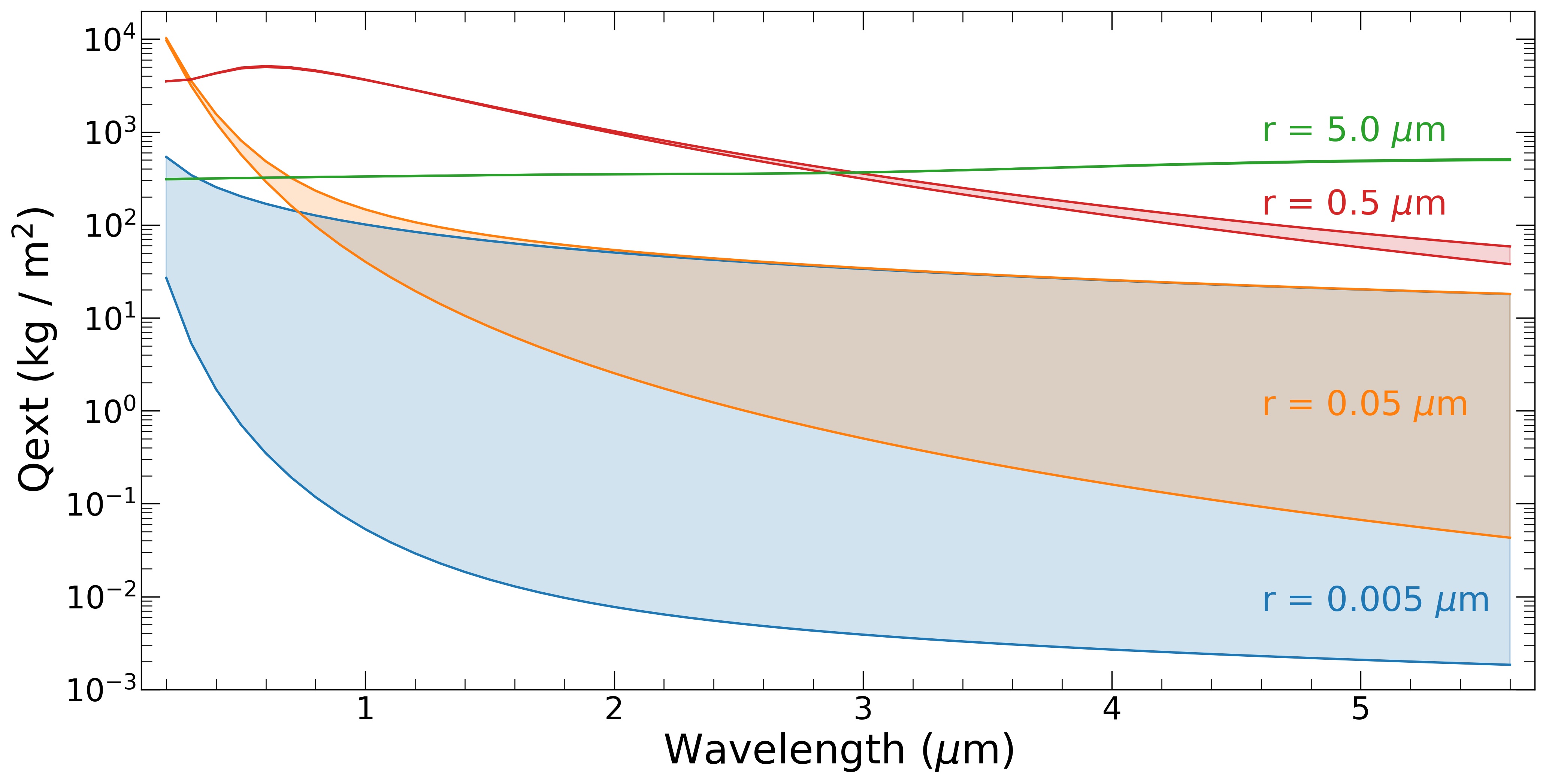}
   \caption{Particle extinction as a function of wavelength for a few particle sizes (assumed spherical) and extreme values of imaginary refractive index $n_i = 10^{-2}$ and $n_i = 10^{-6}$ (solid color lines). The intermediate color areas represent all the possible values that extinction curve may take within the given limits.}
              \label{FigEffectNi}
    \end{figure}

One final question would be whether or not an imaginary refractive index depending on wavelength would provide a good fit to the data. It must be noted that, in most solar system atmospheres, the observed particles in the upper atmosphere are still to be identified and show unexpected absorptions (e.g. \citealt{Toledo2019, Arteaga2023}). As computation time makes this unfeasible for a complete exploration of the free parameter space, we show in Figure \ref{FigEffectNi} a preliminary exploration of this problem. We use upper and lower limits for the particle absorption for a number of particle sizes and test how much the particle extinction may change. We find that very small particles show enormous differences depending on the actual values of the particle absorption, but particles with sizes around observation wavelengths do not. This implies that composition, and hence $n_i(\lambda)$, would be of little relevance in transit geometry for the particle sizes obtained in this work, provided that they can be treated as Mie scatterers.

\section{Conclusion}\label{sec:Conclusions}

JWST observations have an enormous potential to start studying in depth the characteristics of clouds on exoplanets. The wider spectral range provided by instruments onboard this telescope perfectly matches the smooth and extended extinction spectral footprint that can be expected from more realistic prescription of the aerosols. It has been demonstrated in this work how such more realistic prescriptions of the aerosols are strongly supported by the data and fit much better than other simpler models that have been usually assumed as first educated guess for the problem.

In particular, we found two different aerosol spectral behaviours being statistically favoured over a flat contribution. A heavy decreasing opacity with wavelength, resembling the effect of small particles, and a slight increase of opacity resembling the effect of larger aerosol particles. Nevertheless, after our last calculations, the increased retrieved mean molecular weight and the constrain of methane abundance despite the fact that it has not been detected in any previous works (e.g. \citealp{Rustamkulov2023} or \citealp{Lueber2024}) has led us to consider that Model III is not appropriate for defining the aerosol contribution when studying the atmosphere of Wasp-39b. Thus, we conclude that a cloud extinction model with an opacity slightly growing with wavelength is the most appropriate one. Specifically, Mie scattering from a particle distribution of effective radius around 3 $\mu$m is our best guess for fitting the observational data.

It has been shown that there is a significant bias between the retrieved atmospheric parameters obtained with more complex particle simulations and with the simpler flat model. It is important then to implement aerosol realistic simulations not only for getting a better understanding of clouds on exoplanets but also for avoiding any bias in the rest of parameters under study. 

Nevertheless, we have also found some limitations to the degree of complexity that we can include in the aerosol spectral description. We found for example not much sensitivity to the particle absorption, which would probably require a geometry different from the transit observations. Other particle properties, such as single-scattering albedos or scattering asymmetry factors, can be constrained using reflected light phase curves \citep{Heng2021, Morris2024}.

\begin{acknowledgements}
      This work was supported by Grupos Gobierno Vasco IT1742-22. It has also been supported by grant  PID2023‐149055NB‐C31 funded by MICIU/AEI/10.13039/501100011033 and FEDER, UE. J. Roy-Perez acknowledges a PhD scholarship from UPV/EHU. The authors acknowledge the effort of the scientists involved in JWST ERS programs and their contribution to the community. We thank Dr. G. Villanueva and the PSG team for the development of the tool and the support provided to the users.
      
\end{acknowledgements}

\bibliographystyle{aa}
\bibliography{aanda}

\begin{appendix}

\onecolumn

\section{Additional figures}

\begin{figure}[h!]
   \centering
   \includegraphics[width=\hsize]{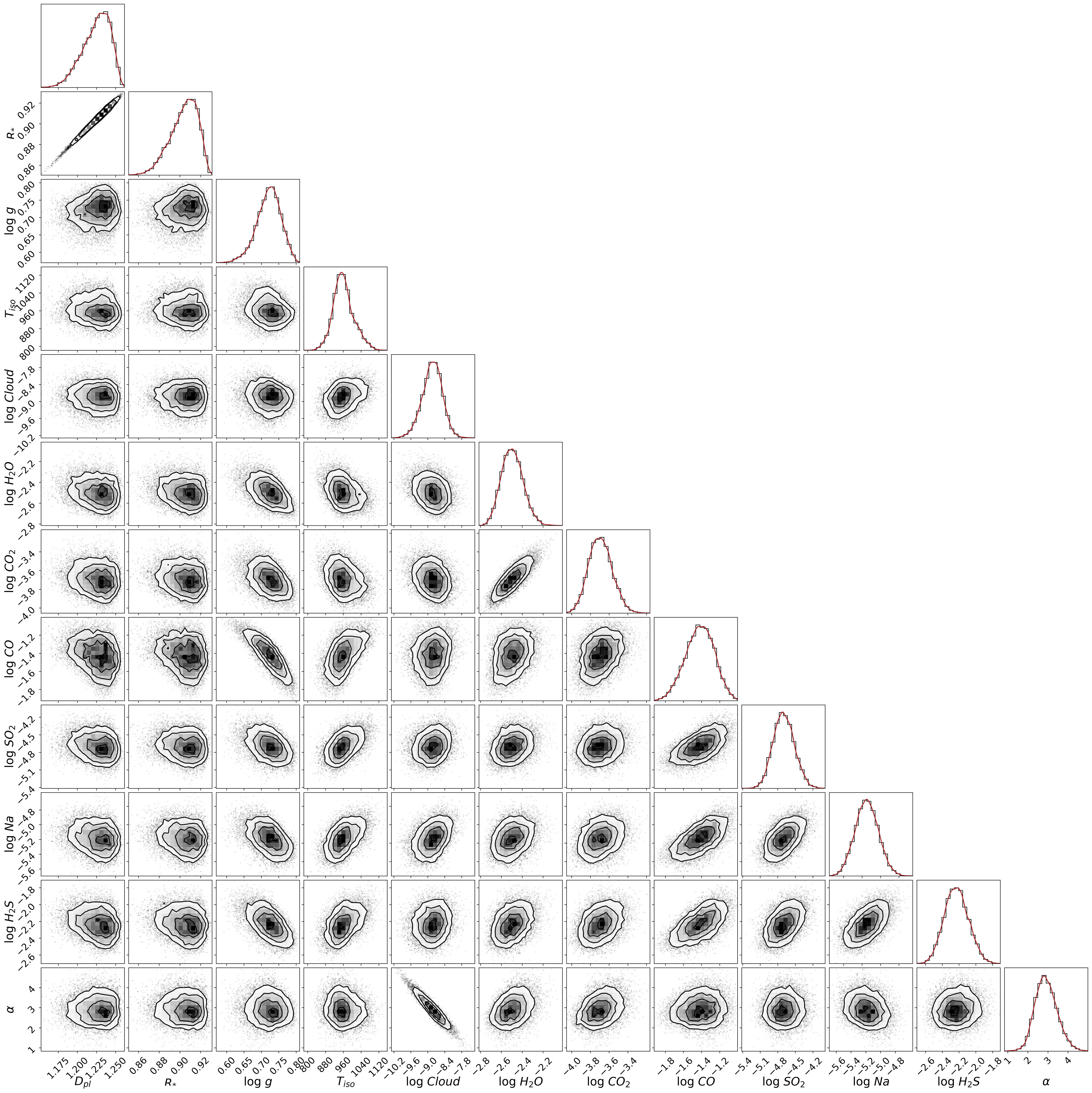}
   \caption{Corner plot of the retrieval using the $\textup{\r{A}}$ngström cloud extinction model (model III). Top plots for each column are the marginalised probability distribution for the different free parameters of the retrieval. Red lines are the fit of those distributions to a Gaussian function and vertical dashed lines are the median and the 1 $\sigma$ deviations. The rest of plots present the marginalised probability distributions as a function of pair of parameters. Parameters have been represented in units shown in Table \ref{t4}. $D_{pl}$, however, has been represented in units of Jovian diameters.}
              \label{FigCornerPlot}
    \end{figure}

\begin{figure}[h!]
   \centering
   \includegraphics[width=\hsize]{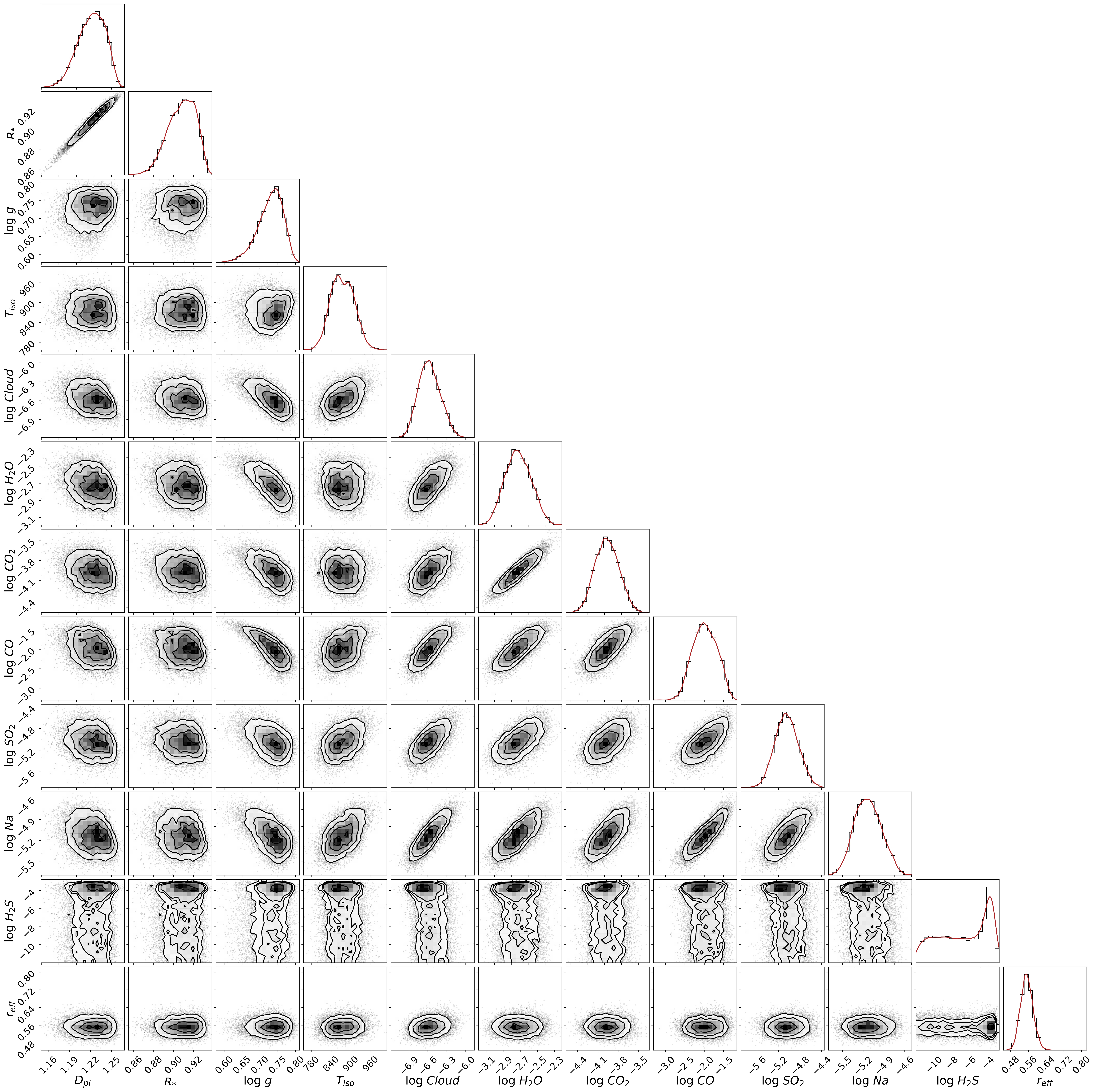}
   \caption{Corner plot of the retrieval using the MOPSMAP cloud extinction model (model IV). Top plots for each column are the marginalised probability distribution for the different free parameters of the retrieval. Red lines are the fit of those distributions to a Gaussian function and vertical dashed lines are the median and the 1 $\sigma$ deviations. The rest of plots present the marginalised probability distributions as a function of pair of parameters. Parameters have been represented in units shown in Table \ref{t4}. $D_{pl}$, however, has been represented in units of Jovian diameters.}
              \label{FigCornerPlot2}
    \end{figure}

\end{appendix}

\pdfoutput=1

\end{document}